\documentclass[conference]{IEEEtran}
\IEEEoverridecommandlockouts

\usepackage[T1]{fontenc}
\usepackage{tabularx}
\usepackage{cite}
\usepackage{graphicx}
\usepackage{amsmath}
\usepackage{amsmath}
\usepackage[cmintegrals]{newtxmath}
\usepackage{algorithmic}
\usepackage{caption}
\usepackage{subcaption}
\usepackage{xcolor}
\usepackage[para]{footmisc}
\usepackage{tabularx,multirow,array}
\usepackage{multirow}
\usepackage{booktabs}
\usepackage[linesnumbered,ruled,vlined]{algorithm2e}
\SetKwInput{KwInput}{Input}                
\SetKwInput{KwOutput}{Output}  
\usepackage{balance}

\begin{document}
\title{Reinforcement Learning-based Dynamic Service Placement in Vehicular Networks}

\author{\IEEEauthorblockN{Anum Talpur and Mohan Gurusamy}
	\IEEEauthorblockA{\textit{Department of Electrical \& Computer Engineering} \\
		\textit{National University of Singapore, Singapore}\\
		Email: anum.talpur@u.nus.edu, gmohan@nus.edu.sg}
}

\maketitle

\begin{abstract}
The emergence of technologies such as 5G and mobile edge computing has enabled provisioning of different types of services with different resource and service requirements to the vehicles in a vehicular network. The growing complexity of traffic mobility patterns and dynamics in the requests for different types of services has made service placement a challenging task. A typical static placement solution is not effective as it does not consider the traffic mobility and service dynamics. In this paper, we propose a reinforcement learning-based dynamic (RL-Dynamic) service placement framework to find the optimal placement of services at the edge servers while considering the vehicle’s mobility and dynamics in the requests for different types of services. We use SUMO and MATLAB to carry out simulation experiments. In our learning framework, for the decision module, we consider two alternative objective functions - minimizing delay and minimizing edge server utilization. We developed an ILP based problem formulation for the two objective functions. The experimental results show that 1) compared to static service placement, RL-based dynamic service placement achieves fair utilization of edge server resources and low service delay; and 2) compared to delay-optimized placement, server utilization optimized placement utilizes resources more effectively, achieving higher fairness with lower edge-server utilization.
\end{abstract}


\IEEEpeerreviewmaketitle

\section{Introduction}
\label{sec:intro}
\IEEEPARstart{T}{he} platform of fifth-generation network (5G) brings tremendous benefits in vehicular communications, including transportation efficiency, improved safety, high reliability, low latency, and large communication coverage. 5G networks are highly-flexible and programmable end-to-end networks that provide enhanced performance while meeting various requirements from multiple services. The International Telecommunications Union (ITU) has categorized the diverse 5G services into three major use-cases, namely, (i) enhanced Mobile Broadband (eMBB) supporting very high data rate of $\approx$10 Gbps, (ii) ultra-Reliable and Low Latency Communications (URLLC) with high reliability and very low end-to-end latency of about 1 ms, and, (iii) massive Machine Type Communications (mMTC) supporting a density of $\approx$106 devices/km2 \cite{ITUSlice}. To support vertical applications of different performance requirements, the next-generation mobile network (NGMN) alliance has introduced the concept of network slicing\cite{slicing}. Network slices are virtual entities (or virtual network functions) that are deployed on a common-physical infrastructure to satisfy the diverse requirements of the use-cases in terms of functionalities and performance. A mobile operator is allowed deploy different slices in parallel while guaranteeing isolation so that services of one slice do not affect services in another slice. \par
\begin{figure}[htbp]
	\begin{center}
		\includegraphics[width=3.1in,height=1.55in]{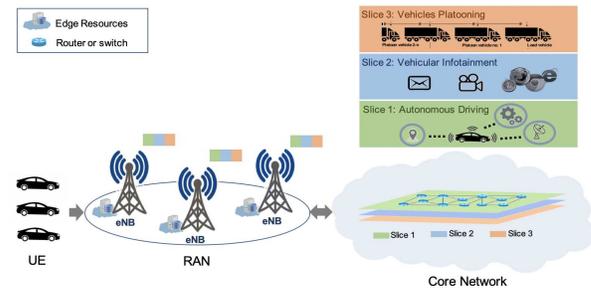}
		\caption{An example 5G slicing system for vehicular applications}
		\label{fig:5GModel}
	\end{center}
\end{figure}

In a 5G-based vehicular network, a slice could be instantiated for a specific application, for which each slice is capable of providing multiple services. Vehicles communicate with each other or surrounding infrastructure for availing coordination in driving, convenience, road safety, and many other applications. Fig. \ref{fig:5GModel} shows an example model of a 5G slicing system for three different vehicular applications including, autonomous driving, vehicular infotainment, and vehicle platooning. To support these applications, several computational operations need to be performed within a network. The European Telecommunications Standards Institute (ETSI) outlines the use of mobile edge computing (MEC) along with vehicular networks to satisfy various vehicle-to-everything (V2X) service requirements including, low delay, low computational cost, security, reliability, and so on\cite{etsiMEC}. The closer the application server is to a vehicle, the faster the communication and better the coverage is for the vehicle\cite{MECSurvey}. \par 
As shown in Fig. \ref{fig:5GModel}, the 5G slices support three kinds of applications in a vehicular network. The slices share the resources of radio access network (RAN), edge and core networks. Services can be deployed at the servers making use of compute, network and storage resources.  Service placement is the problem of mapping services to the edge servers in vehicular networks to satisfy the requirements for the requested services while utilizing the resources efficiently. From the perspective of vehicles, the delay perceived by a vehicle is an important metric which should be minimum. From the service provider perspective, it is important to minimize the server resource utilization while  keeping the resource utilization across the servers as balanced as possible. This will enable servers to scale up the resources to handle the events of congestion, failures and changing service demands.  \par
The growing complexity of traffic patterns and dynamics in the requests for different types of services has made service placement more challenging. It is necessary to adopt continual learning of the environment for providing better services. Therefore, embedding intelligence using machine learning (ML) has drawn research interests recently. Different ML techniques have received significant attention and reinforcement learning (RL) is an attractive approach for various problems in the area of vehicular communications \cite{RLforITS}. \par 
In \cite{Edge-enabledV2X}, the authors addresses the problem of V2X service placement using delay as the objective function. It performs fixed placement based on the optimization formulation considering a static service placement problem that does not take into account the dynamicity of service requests. A static solution in \cite{Edge-enabledV2X} that fixes servers for hosting services is not effective for a mobile and dynamic scenario of a vehicular network. It is therefore imperative that the real-time environment be taken into consideration while mapping service to an edge server. While our work considers a  delay-based optimization similar to \cite{Edge-enabledV2X} , we use it as one of the two possible objective functions to be used by the decision module in our learning framework. Different from \cite{Edge-enabledV2X}, we consider the dynamic service placement problem with the proposal of a new solution approach based on reinforcement learning to continuously learn and adapt to the dynamics of the system. \par 
In this paper, we propose a reinforcement learning-based dynamic (RL-Dynamic) service placement framework to find the optimal placement of services at the edge servers while considering the vehicle's mobility and dynamics in the requests for different types of services. We use SUMO and MATLAB to carry out simulation experiments. In our learning framework, for the decision module, we consider two alternative objective functions - minimizing delay and minimizing edge server utilization. We developed an ILP based problem formulation for the two objective functions. The experimental results show that 1) compared to static service placement, RL-based dynamic service placement achieves fair utilization of edge server resources and low service delay; and 2) compared to delay-optimized placement, server utilization-optimized placement utilizes resources more effectively, achieving higher fairness with lower average edge-server utilization. \par
The remaining sections are organized as follows. Section \ref{sec:relatedwork} provides an overview of the related work in the literature. Section \ref{Sec:System-Model} decribes the network model, request model and problem description. Section \ref{sec_proposedmethod}  presents the proposed method. Section \ref{sec:results} discusses the experimental setup and results, and section \ref{sec:conclude} concludes the paper.

\section{Related Work}
\label{sec:relatedwork}
In the literature, there are few recent works focusing on the problem of optimal service placement for vehicular networks. In \cite{Edge-enabledV2X}, the authors consider the problem of V2X service placement. They propose an ILP model for minimizing the average service delay where the scope of the environment is limited to the highway with constant speed, fixed distance between vehicles, and movement of vehicles in one direction. The delay experienced by the vehicles for V2X communication is also based on randomly assigned values from a given range. In \cite{multicomponent-V2X}, the authors consider a more realistic scenario of the highway environment for V2X service placement. They consider 5 different V2X applications for minimization of communication and download link delay using binary ILP model. Some work on V2X applications are carried out in the context of cloud computing and fog-computing \cite{cloud1,cloud2,cloud3,priority1}. One common aspect in most of the previous works is the consideration of latency or delay as the objective. Some works consider priority \cite{priority1} and cost \cite{cloud2,cloud3} as additional factors for service migration. The above works basically consider a static service placement problem which do not consider the dynamicity of service requests. Different from the existing works, our work makes contributions in consideration of the dynamic service placement problem and development of a reinforcement learning-based dynamic (RL-Dynamic) solution to provide services with low delay while keeping the server resource utilization low. In our learning framework, for the decision module, we consider two alternative objective functions - minimizing delay and minimizing edge server utilization. 

\section{Problem Description}
\label{Sec:System-Model}
In this section, we describe the problem and system model of the proposed approach.

\subsection{Assumptions}
In this paper, we assume a city road environment with multiple lanes in different directions. The vehicles are moving along the road by randomly choosing a source, destination, and speed to start and end their journey at different times. The speed limit regulations specified in the SUMO simulator, for the type of vehicle and environment are followed. We assume the city environment is under 5G coverage using evolved NodeB (eNB) stations. 5G is a perfect enabler for V2X applications \cite{5GRAN}. The deployment of eNB follows urban-macro 5G regulations for which inter-site distance (ISD) is 500m \cite{etsi5G}. It is also assumed that eNBs are equipped with MEC hosts to form the network edge with limited capacity servers. Additionally, the network edge connects the core cloud data center with large capacity servers via a backbone network. Further, we assume adequate links between different nodes and servers to enable the communication. \par
\subsection{Network Model}
We use $E$ to denote a set of edge servers with $i$ $\epsilon $ $E$ as an edge node. For each edge node $i$, the residual computing resources (available resources) is denoted by $C_i$. Let $V$ and $S$ denote a set of vehicles (UEs) and service types (services), respectively. A vehicle $\nu$ $\epsilon $ $V$ requires a service $s$ $\epsilon $ $S$ which is to be hosted at an edge node. The amount of resources consumed by deploying service $s$ at edge node $E$ is denoted by $R_s$, and the delay/latency requirement for each service $s$ is denoted as $D_s$. The notations are summarized in Table. \ref{tab:notations}. \par
\begin{table}[htbp]
   \scriptsize
  \centering
  \caption{Summary of Notations}
    \begin{tabular}{lp{6.5cm}}
    \toprule
    \textbf{Notation} & \textbf{Description} \\
    \midrule
    $E$    & Set of edge servers \\
    $S$     & Set of services \\
    $V$     & Set of vehicles \\
    $R_s$    & Resources consumed by service $s$ \\
    $C_i$ & Available resources at edge node $i$ \\
	$x_i^s$ & Assignment of service $s$ at the edge node $i$ \\
    $\varphi_i$    & Server utilization of edge node $i$ \\
    $\beta$  & Balancing factor \\
    $d^s_{i,v}$  & The time delay experienced by vehicle $v$ when services $s$ is deployed at node $i$ \\
    $D_s$ & Delay threshold or maximum allowed delay for service $s$  \\
    $U_s$ & Service demand (number of UEs requesting service $s$) \\
    $N_i$    & Number of UEs edge node can handle \\

    \bottomrule
    \end{tabular}%
  \label{tab:notations}%
\end{table}%

\subsection{Service Request Model}
A service request is specified as a 4-tuple ($\nu$, $loc$, $t$, $s$), wherein $\nu$, $loc$, $t$,and $s$ represent the vehicle ID, vehicle location, time of request and type of service, respectively. We assume each entity (i.e. vehicle/UE) is equipped with a clock and GPS, which enables it to specify time $t$ and location $loc$ in its service request message. Associated with each service $s$, there are delay and resource requirements. In response to the request, the location of the edge server where the requested service is deployed, will be returned. 

\subsection{Problem and Proposed Approach}
We consider a service placement problem in vehicular networks with edge servers having limited resources. Given a set of services belonging to different network slices with their resource and delay requirements, the problem is to find the optimal placement of services at the edge servers while considering the vehicle's mobility and dynamics in the requests for different types of services. The number of vehicles requesting service $s$ and their distance from different edge servers is very dynamic. A static solution which fixes servers for hosting services is not effective for a mobile and dynamic scenario of a vehicular network. It is therefore imperative that the real-time environment be taken into consideration while mapping a service to an edge server. With this goal, we propose a RL-based approach and  develop an RL-Dynamic service placement algorithm which facilitates remapping of services to edge servers in accordance with the changing vehicular environment. Our solution framework uses a classic model-free Q-learning algorithm which finds an optimal set of actions $a$ which optimize a certain objective such as minimizing resource utilization or minimizing the delay while satisfying the service requirements.

\section{Proposed RL-based Dynamic (RL-Dynamic) Service Placement Framework}
\label{sec_proposedmethod}
In this section, we present the proposed RL-Dynamic service placement framework, for the problem described above. \par
\begin{figure}[htbp]
	\begin{center}
		\includegraphics[width=3.2in,height=1.55in]{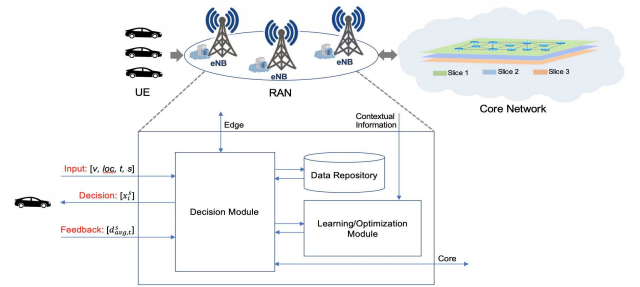}
		\caption{The proposed RL-based dynamic (RL-Dynamic) service placement framework}
		\label{fig:RLModel}
	\end{center}
\end{figure}

 
We first summarize the state space, reward function, the action space and state transition policy used in our RL framework. 
\begin{itemize}
	\item \textit{State Space $\omega_t(s)$}: The state space set describes the network environment, formed from the request messages ([$\nu_1$, $loc_1$, $t$, $s$], [$\nu_2$, $loc_2$, $t$, $s$], ... , [$\nu_n$, $loc_n$, $t$, $s$]) for service $s$ at time $t$, where $n$ is number of vehicles requesting for service $s$. 
	\item \textit{Reward Function $r(\omega_t(s),a_t)$}: The reward is measured from the average delay feedback from vehicles in accessing their service from the associated edge server.
	\item \textit{Action Space and State Transition Policy $a_t$}: The action space describes the action taken by the decision module for placement of service $s$ on the edge node $i$. On the other hand, the state transition policy defines the condition, upon which re-optimization of the decision matrix for the new service placement will take place. It aims to maximize the q-value for the action space. 
\end{itemize}
Fig. \ref{fig:RLModel} depicts the framework of the RL-Dynamic service placement algorithm which consists of three modules (decision module, learning module, and data repository). The \textit{decision module} has direct interaction with the UEs. The request for service $s$ by UE $\nu$ is specified in the form of a 4-tuple ($\nu$, $loc$, $t$, $s$), as discussed in Section \ref{Sec:System-Model}. In return, considering the demand for service $s$ at time $t$ and location $loc$ of vehicles (UEs) requesting for service $s$, the decision module selects the servers for the services to place based on the Q-table maintained in the data repository by the learning module. The \textit{learning module} optimizes the objective function subject to different constraints. We use two alternative objective functions in the framework in our study. First, we consider delay/latency as the objective function for providing a RL-Dynamic solution. Second, we consider edge-server utilization as the objective function. \par
\textit{Delay}: In this work, the delay problem is formulated as, \\
\hspace*{10mm}Minimize
\begin{equation}
\sum_{s\epsilon S_u}\sum_{i\epsilon E}(\frac{1}{|V|}\sum_{v\epsilon V}d_{i,v}^s)x_i^s
\label{eq1}
\end{equation}
\hspace*{10mm}Subject to:
\begin{equation}
\sum_{i\epsilon E}x_i^s = 1; \forall s\epsilon S
\label{eqC1}
\end{equation}
\begin{equation}
\sum_{s\epsilon S_u}x_i^s \le 1; \forall i\epsilon E
\label{eqC2}
\end{equation}
\begin{equation}
\sum_{s\epsilon S_u}x_i^s(\frac{1}{|V|}\sum_{v\epsilon V}d_{i,v}^s) \le D_s; \forall i\epsilon E
\label{eqC3}
\end{equation}
\begin{equation}
\sum_{i\epsilon E}x_i^sR_s \le C_i; \forall s\epsilon S
\label{eqC4}
\end{equation}
\begin{equation}
\sum_{i\epsilon E}x_i^sU_s \le N_i; \forall s\epsilon S
\label{eqC5}
\end{equation}
\begin{equation}
x_i^s \epsilon \{0,1\}; \forall s\epsilon S, \forall i\epsilon E
\label{eqx}
\end{equation}
$x_i^s$ is a binary variable used to indicate the placement of service $s$ at node $i$. If edge node $i$ deploys service $s$, $x_i^s$ is 1. Otherwise, it is 0. Constraint (\ref{eqC1}) ensures that each service should be deployed onto exactly one edge server. Constraint (\ref{eqC2}) guarantees each edge server node hosts a different and unique service. This condition ensures no repetition of the same service at different edge servers. Constraint (\ref{eqC3}) ensures that the average delay experienced by vehicles requesting service $s$ should be less than the service's maximum delay threshold. Constraint (\ref{eqC4}) ensures that the available resources at the edge node is not exhausted while deploying service $s$. Constraint (\ref{eqC5}) guarantees the edge node has the capacity to handle the number of UEs requesting service $s$. Finally, condition (\ref{eqx}) defines the decision variable $x_i^s$ as a binary integer decision variable. \par
\textit{Server-Utilization}: In this case, we consider minimizing the total server resource utilization. The objective function is formulated as, \\
\hspace*{10mm}Minimize
\begin{equation}
\sum_{s\epsilon S}\sum_{i\epsilon E} \varphi_i (U_s + \beta)x_i^s
\label{eq2}
\end{equation}
The rationale for minimizing the server utilization is to decrease the possibility of congestion so that a server has enough room for resource scale-up. Our proposed objective function aims to minimize the total edge server utilization \footnote{Server utilization is defined as the ratio between the resources that any service type $s$ will consume and the available resources at the edge node} weighted by service demand\footnote{Service demand is defined as the total number of UEs requesting for service $s$}. In our objective function, we use a small offset value $\beta$ (i.e. balancing factor) to account for a situation of zero requests for any service as it would otherwise result in multiplication by zero in the objective function. The difference between the two ILP formulations is in the objective function. On the other hand, the constraints are the same. \par 
The decision matrix calculated by the learning module is tabulated for a quick reference by the decision module. It is stored in the \textit{data repository} in the form of a Q-table. A high Q-value means a high-quality decision. The decision module will look-up the Q-table to forward the new decision to a vehicle (UE), and updates the Q-value and reward based on the feedback for the previous decision. In our model, the iterative maintenance of the lookup table (i.e. Q-table) uses the Bellman equation to observe a given state of the environment. The Q-value of a state at a given time is calculated as,
\begin{equation}
\begin{split}
Q^{new}(\omega_t(s),a_t) &=\alpha (r(\omega_t(s),a_t))+ \\
&\gamma (maxQ(\omega_{t+1}(s),a))+(1-\alpha) Q(\omega_t(s),a_t))
\end{split}
\label{eq:Qvalue}
\end{equation}
where, $(\omega_t(s),a_t)$ is a state-action pair at time $t$ and $r(\omega_t(s),a_t)$ is the reward of applying action $a_t$ at state $\omega_t(s)$. The value $\alpha$ and $\gamma$ is the learning rate and discount factor, respectively. The greater the discount factor $\gamma$ is, the more the effect of future reward is applied to the new Q-value. In our case, $\omega_{t+1}(s)$ is not deterministic and we only consider the historic performance and current feedback/reward to update Q-table, therefore we set $\gamma$ $=$ $0$. The larger value of $\alpha$ implies more significance of the current feedback reward. The decision module continuously receives feedback from vehicles about average delay $d^s_{avg,t}$ experienced by vehicles requesting for service $s$ at time $t$. In this work, we assume the reward is either 1,0.5 or -$\infty$. The reward represents the average delay is either increased (0.5), decreased (1), or violating delay threshold (-$\infty$), as shown in Equation \ref{eq:reward}. Here, $\infty$ means a very large value which converts q-value into a negative value to trigger the violation of constraints. This may vary with the size and scope of a network. In our model, we use the value -10.
\begin{equation}
r(\omega_t(s),a_t) =
\begin{cases}
0.5 & d_{avg,t-1}^s\le d_{avg,t}^s < D_s\\
+1 & d_{avg,t-1}^s\ge d_{avg,t}^s < D_s\\
-\infty & \text{else}
\end{cases} 
\label{eq:reward}
\end{equation}
We present the proposed RL-Dynamic service placement technique in Algorithm \ref{Alg:1}. In line 5-10, the learning module solves ILP to calculate optimal decision matrix $x_i^s$ at time $t=1$ for the given set of requests. The decision module performs action $a_t$ (i.e. forwarding information of corresponding service and its deployment location to the vehicles) for all services. It also stores the decision matrix in the form of a lookup table at the data repository. The iterative re-optimization of the decision matrix will take place after every fixed period to capture the dynamicity of vehicular networks (Line 11-18). The decision module monitors the state of the Q-value for the stored action set. According to the state transition policy used in this work, if the Q-value is decreased twice consequently for any service $s$ or the constraint of delay is violated then the learning module performs re-optimization of $x_i^s$ over the newly-collected set of data at the given time. This is updated in the Q-table as well. The state transition policy may vary depending on the size of a network and the periodicity of monitoring q-values.  
\begin{algorithm}
	\DontPrintSemicolon
	\KwInput{[$v$, $loc$, $t$, $s$]}
	\KwOutput{State-Action pair $(\omega(s),a)$}
	\textbf{Initialization:} $r(\omega_t(s),a_t)=0$, $Q(\omega_t(s),a_t)=0$ \\
	Set $d_{avg,t}^s = \frac{1}{|V|}\sum_{v\epsilon V}d_{i,v}^s$ \\
	\For{t=1,2,3,....}
	{
		\For{all s$\epsilon$S}
		{
			Collect $\omega_t(s)$ \\
			\If{t=1}
			{calculate $x_i^s$ using ILP (learning-module) \\
			 set $Q(\omega_t(s),a_t)=x_i^s$ \\
			 create Q-table from $x_i^s$, $Q(\omega_t(s),a_t)$, and $r(\omega_t(s),a_t)$ \\
			 perform \textit{action} $a_t$
			}
			\Else
			{perform \textit{action} $a_t$ using Q-table \\
			 calculate feedback $d_{avg,t}^s$ and store \\
			 calculate reward $r(\omega_t(s),a_t)$ using (\ref{eq:reward}) \\
			 Update $Q(\omega_t(s),a_t)$ using (\ref{eq:Qvalue}) \\
			 \If{Consecutive\_Decrements($Q(\omega_t(s),a_t)$) == true || Constraint\_Violated == true}
			 {Reoptimize $x_i^s$ using ILP for the new set of data \\
			  Update Q-table
			 } 
			}
		}
	}
	\caption{RL-Dynamic Service Placement}
	\label{Alg:1}
\end{algorithm}

\section{Experimental Setup and Results Discussion}
\label{sec:results}
To evaluate the performance of the proposed RL-Dynamic service placement mechanism, we use SUMO and MATLAB to set up the simulation environment. SUMO is an open-source simulator, used to simulate a virtual traffic scenario of a realistic vehicular network. In this work, we extract the area of $3km^2$ around the National University of Singapore using Openstreetmaps. The choice of the area is significant as it is present in the center of the city with high traffic densities (Urban environment). Furthermore, the randomTrip application of the SUMO package is used to automatically generate the trips for the vehicles with mobility over the given area of the map. We collect traces of data which helps to generate a 4-tuple message dynamically for our algorithm. The implementation of RL-based optimization is carried out using MATLAB. All experiments are evaluated on a system with Intel Corei5 2GHz and 8GB RAM. The implementation of the proposed RL-Dynamic mechanism for a given set of vehicles has a significantly low run time of $\approx$0.01-0.02sec. \par 
Table \ref{tab:sim-parameters} lists the parameters used in the simulation. Different sets of values are chosen for performing multiple experiments. Small delay threshold is chosen to enforce strict delay constraints. Whereas, the selection of resource unit for $R_s$ and $C_i$ is random. Experiments were performed for different sets (by choosing lowest values as well as the highest values) of $R_s$ and a similar performance trends are observed. We use average service delay, percentage of server utilization, and fairness in edge server utilization as our evaluation metrics. We evaluate our algorithm termed as RL-Dynamic for the two objective functions, delay optimization (D-Optimization) and server utilization optimization (SU-Optimization). We compare our algorithms with the static solution (termed Static) which has a fixed placement based on the optimization solution. We note that the varying distance between a vehicle and the server hosting the service due to the mobility has an impact on the delay and service placement. We carry out experiments (trials) five times with different random seeds. We present the results for different trials. We also present the average of five trials to ensure low confidence intervals. 
\begin{table}[htbp]
  \centering
  \caption{Simulation Parameters}
    \begin{tabular}{ll}
    \toprule
    \textbf{Parameters} & \textbf{Value} \\
    \midrule
    $S$     & 6 \\
    $V$     & 100 \\
    $E$    & 6 \\
    $R_s (Unit)$ & [60 20 60 40 50 70] \\
    $C_i (Unit)$ & [60 60 70 80 90 100] \\
    $D_s (ms)$ & [5 4 4.5 5 5 5.5] \\
    $N_i$    & 100 \\
    $\alpha$ & 0.75 \\
    $\beta$  & 0.1 \\
    $\gamma$  & 0 \\
    $t (sec)$  & 1 to 500 \\
    \bottomrule
    \end{tabular}%
  \label{tab:sim-parameters}%
\end{table}%

\subsection{Average Service Delay}

\begin{figure*}[htbp]
	\centering
	\begin{subfigure}{.19\textwidth}
		\centering
		\includegraphics[width=1.55in]{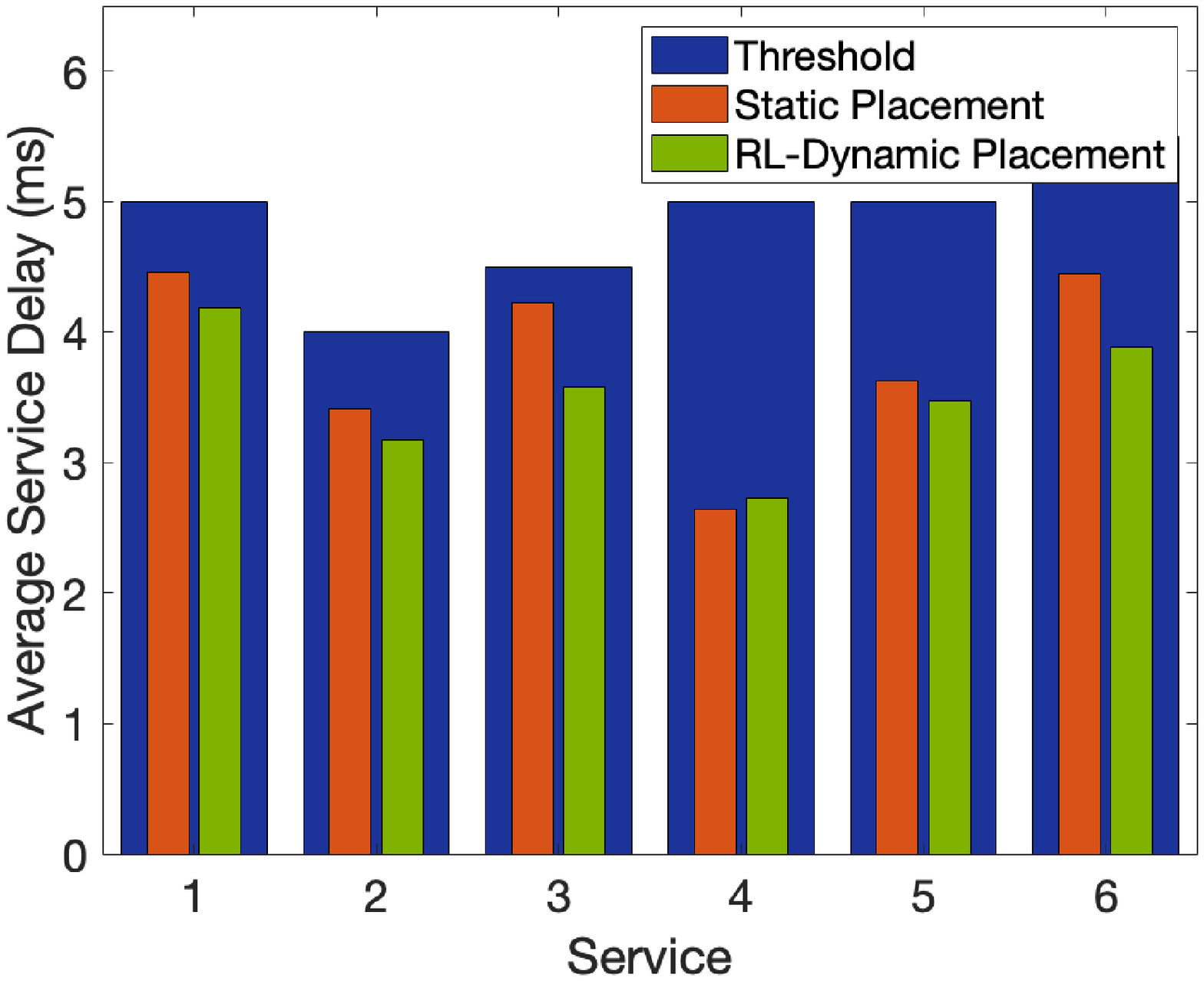}  
		\caption{D-Opt (Trial 01)}
		\label{fig:delay1}
	\end{subfigure}	 
	\begin{subfigure}{.19\textwidth}
		\centering
		\includegraphics[width=1.55in]{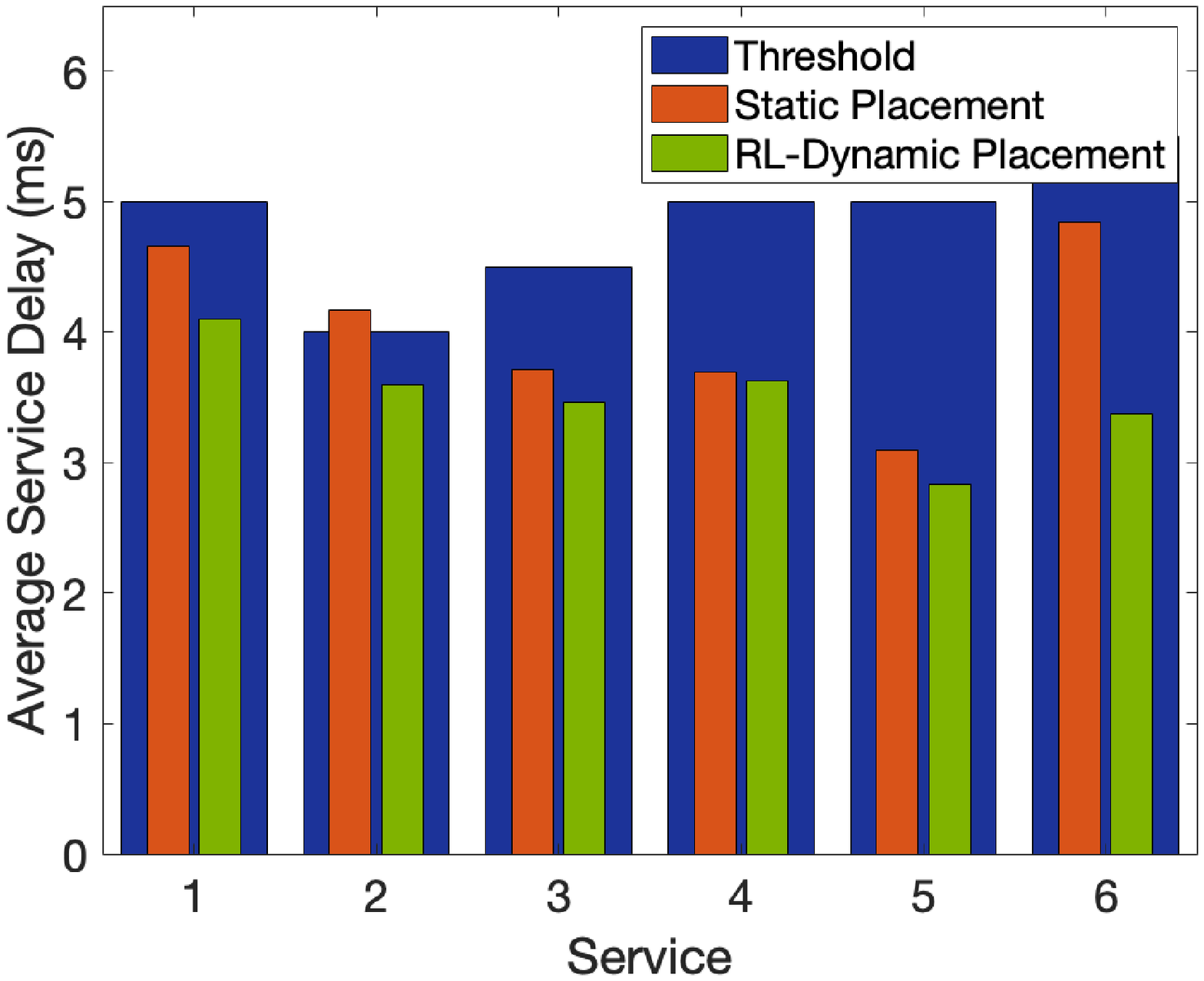}  
		\caption{D-Opt (Trial 02)}
		\label{fig:delay2}
	\end{subfigure}
	\begin{subfigure}{.19\textwidth}
		\centering
		\includegraphics[width=1.55in]{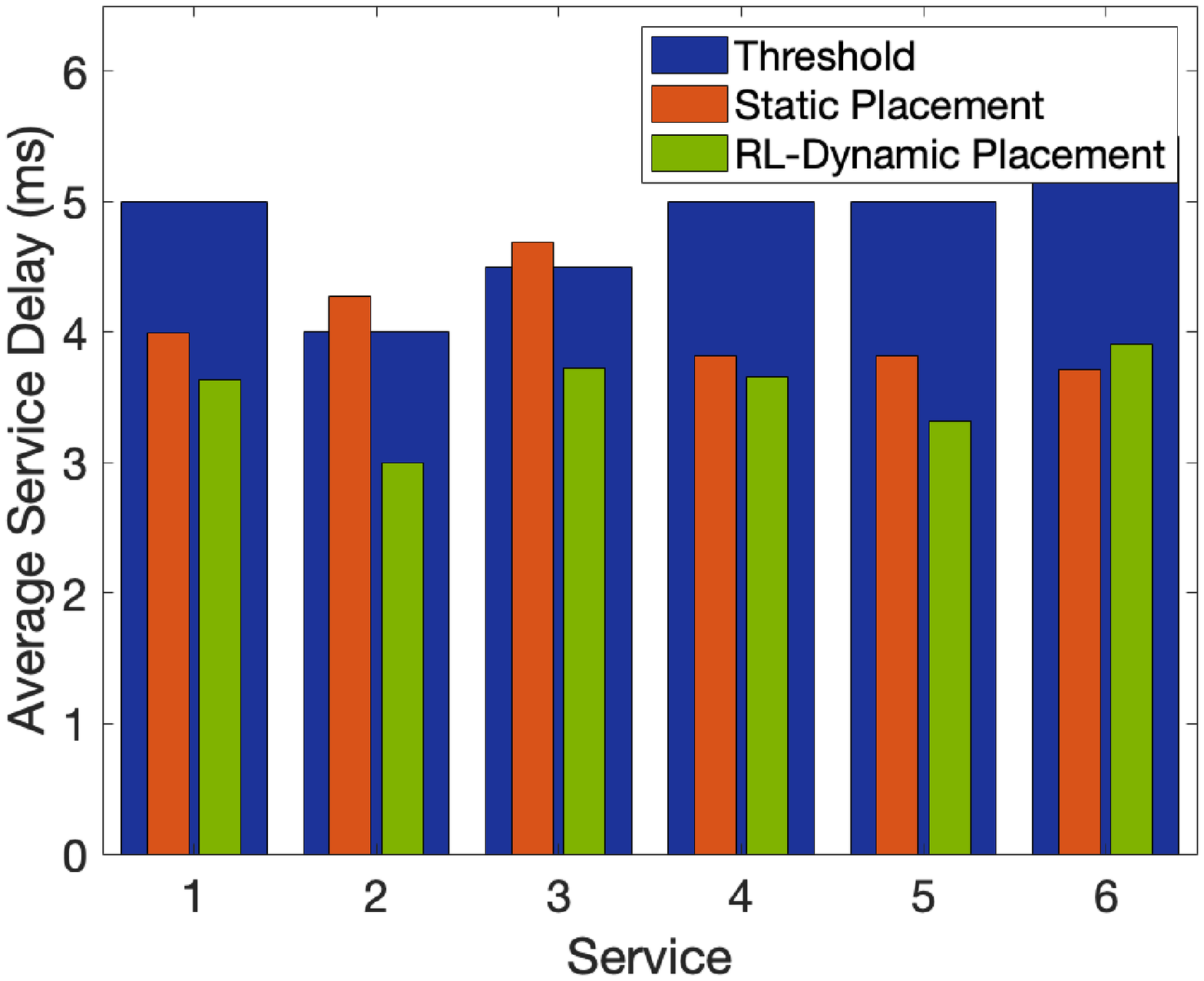}  
		\caption{D-Opt (Trial 03)}
		\label{fig:delay3}
	\end{subfigure} 
	\begin{subfigure}{.19\textwidth}
		\centering
		\includegraphics[width=1.55in]{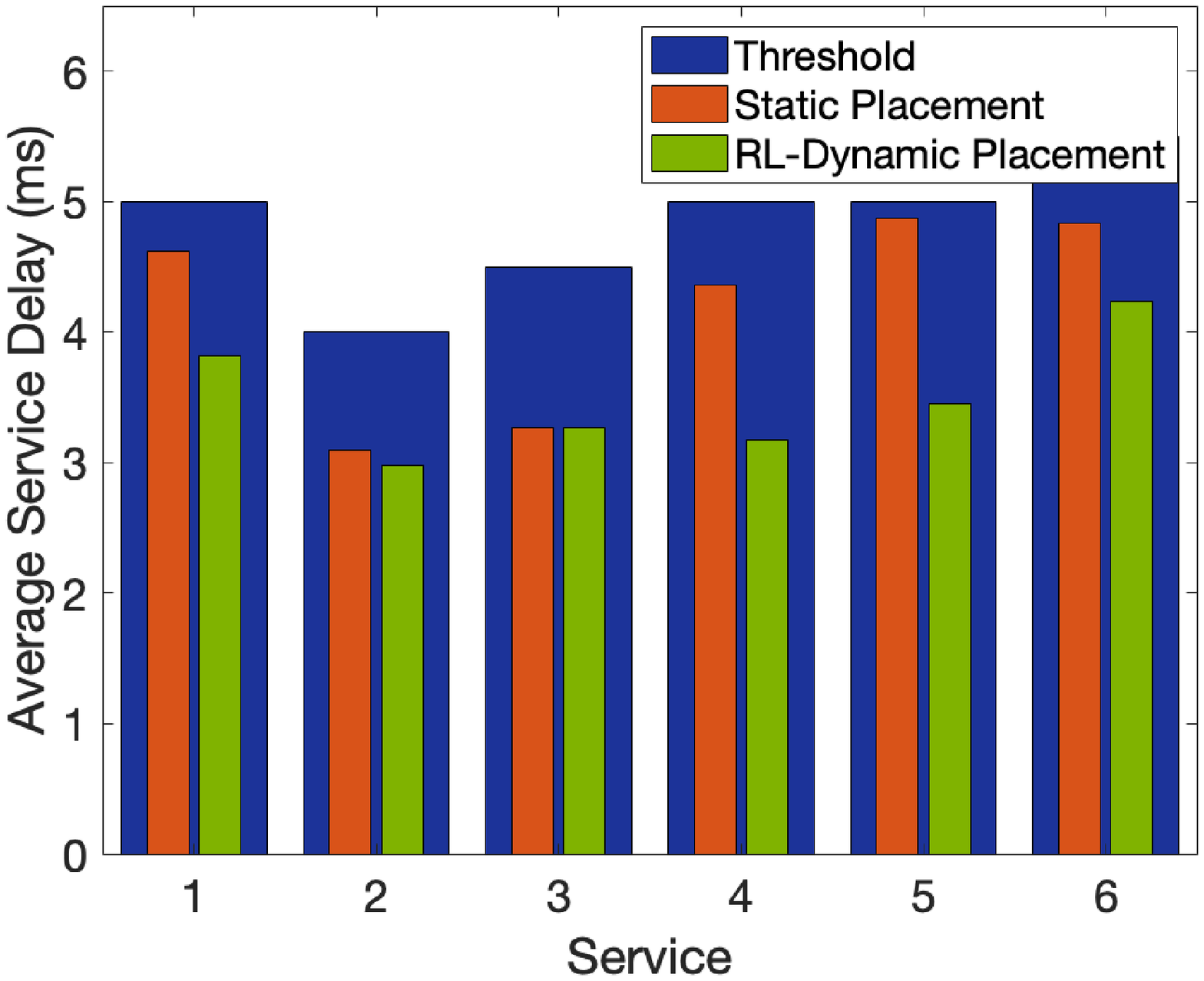}  
		\caption{D-Opt (Trial 04)}
		\label{fig:delay4}
	\end{subfigure} 
	\begin{subfigure}{.19\textwidth}
		\centering
		\includegraphics[width=1.55in]{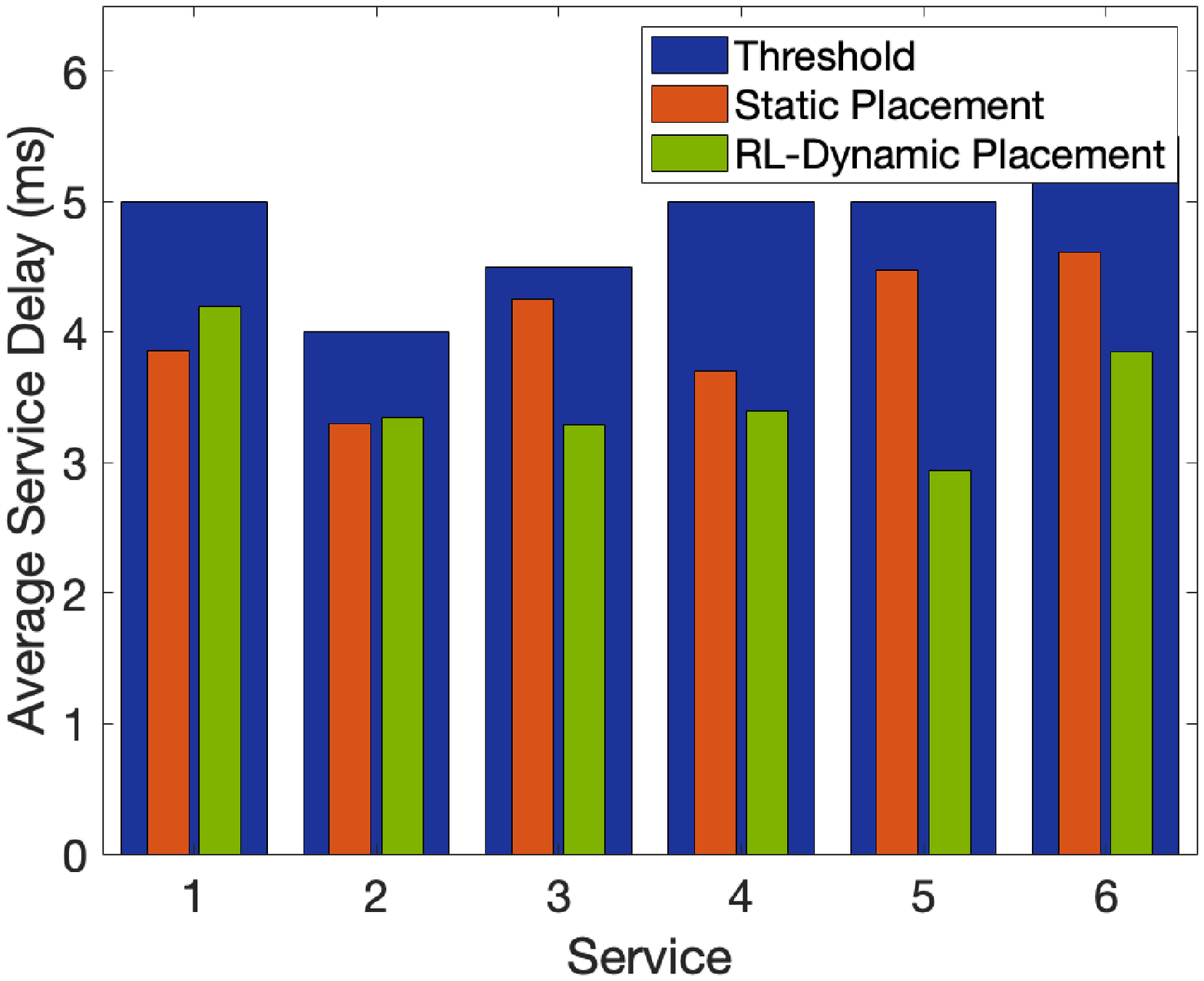}  
		\caption{D-Opt (Trial 05)}
		\label{fig:delay5}
	\end{subfigure} \\
	\begin{subfigure}{.19\textwidth}
		\centering
		\includegraphics[width=1.55in]{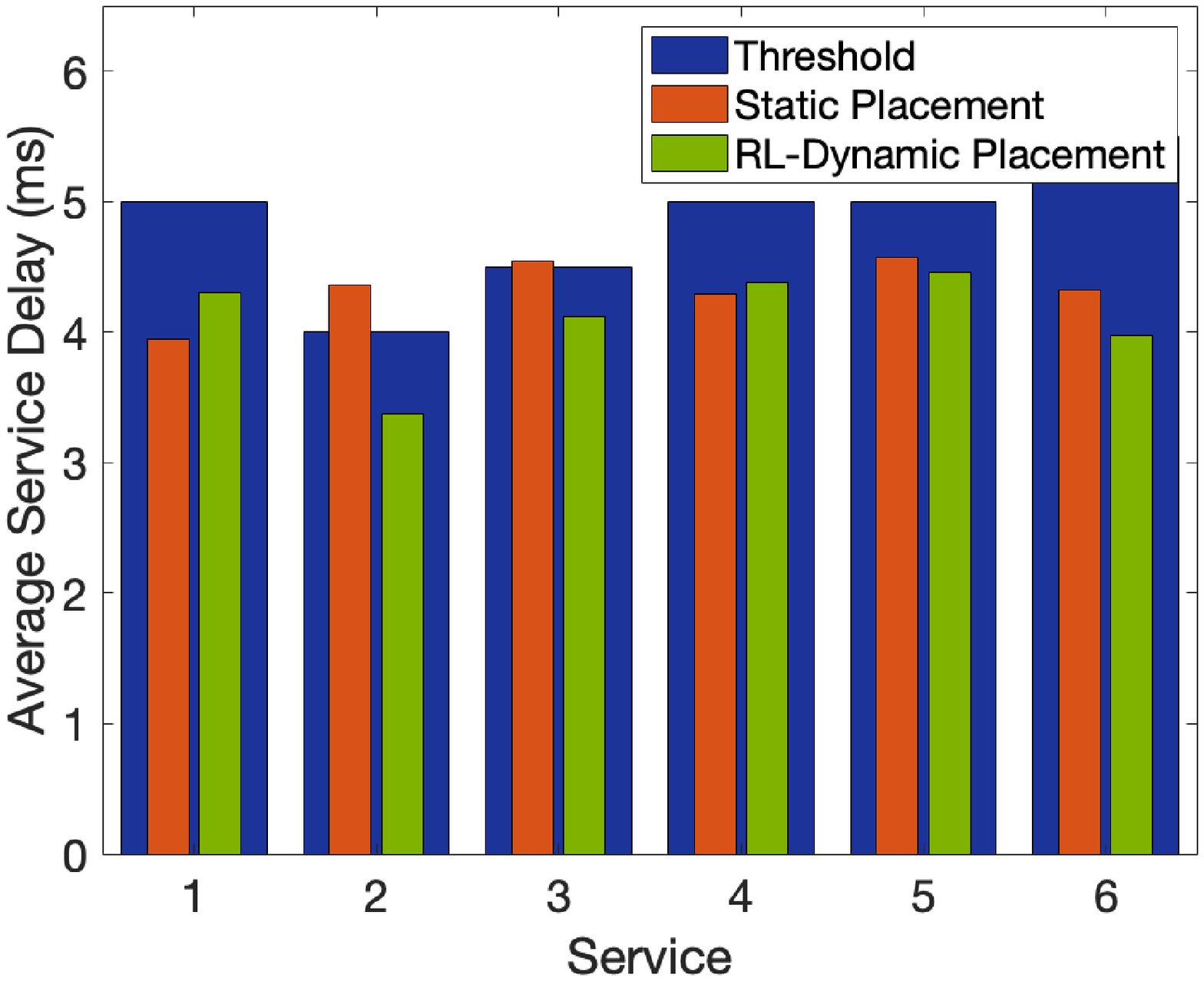}  
		\caption{SU-Opt (Trial 01)}
		\label{fig:delay6}
	\end{subfigure}  
	\begin{subfigure}{.19\textwidth}
		\centering
		\includegraphics[width=1.55in]{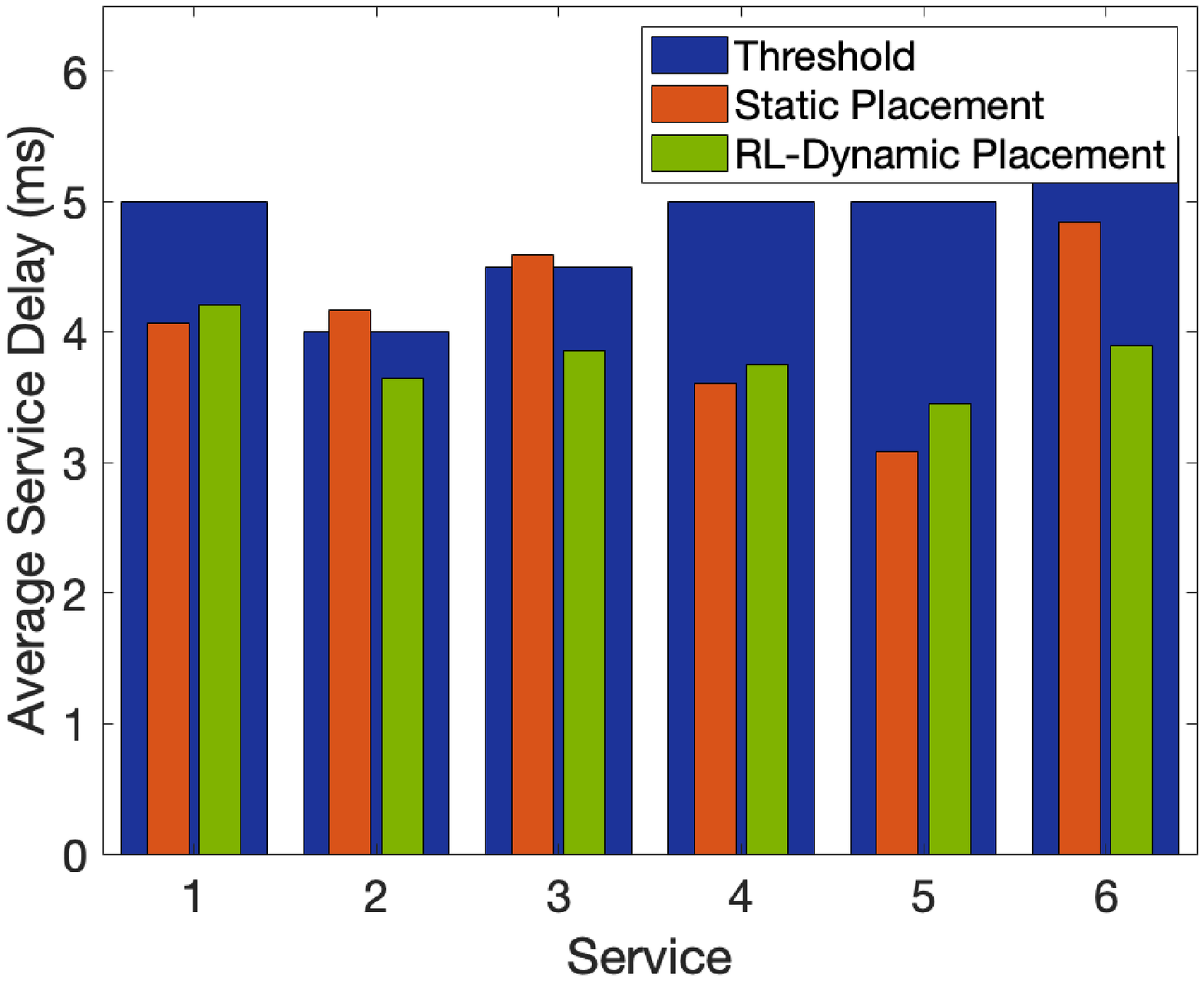}  
		\caption{SU-Opt (Trial 02)}
		\label{fig:delay7}
	\end{subfigure} 
	\begin{subfigure}{.19\textwidth}
		\centering
		\includegraphics[width=1.55in]{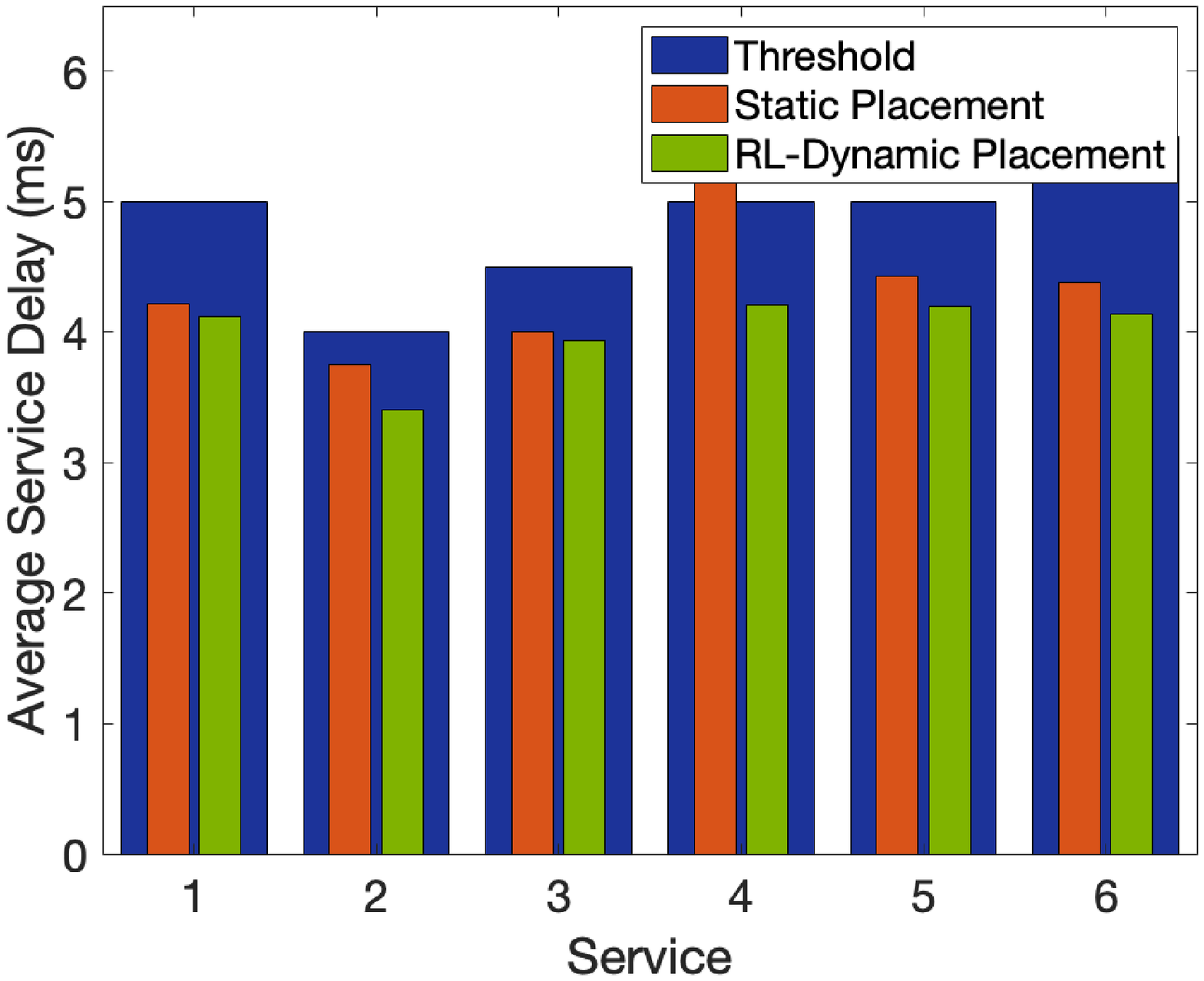}  
		\caption{SU-Opt (Trial 03)}
		\label{fig:delay8}
	\end{subfigure}  	
	\begin{subfigure}{.19\textwidth}
		\centering
		\includegraphics[width=1.55in]{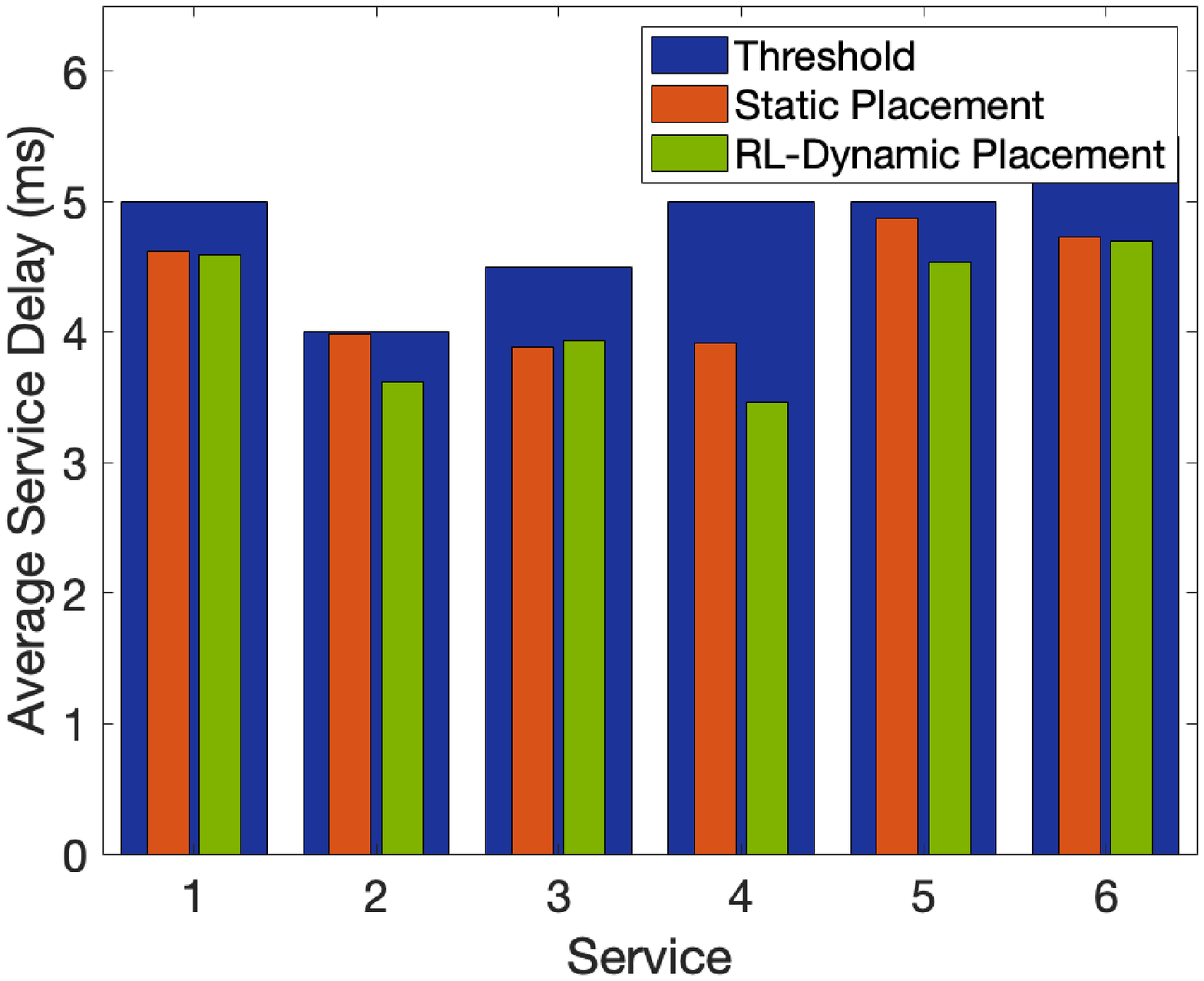}  
		\caption{SU-Opt (Trial 04)}
		\label{fig:delay9}
	\end{subfigure}  
	\begin{subfigure}{.19\textwidth}
		\centering
		\includegraphics[width=1.55in]{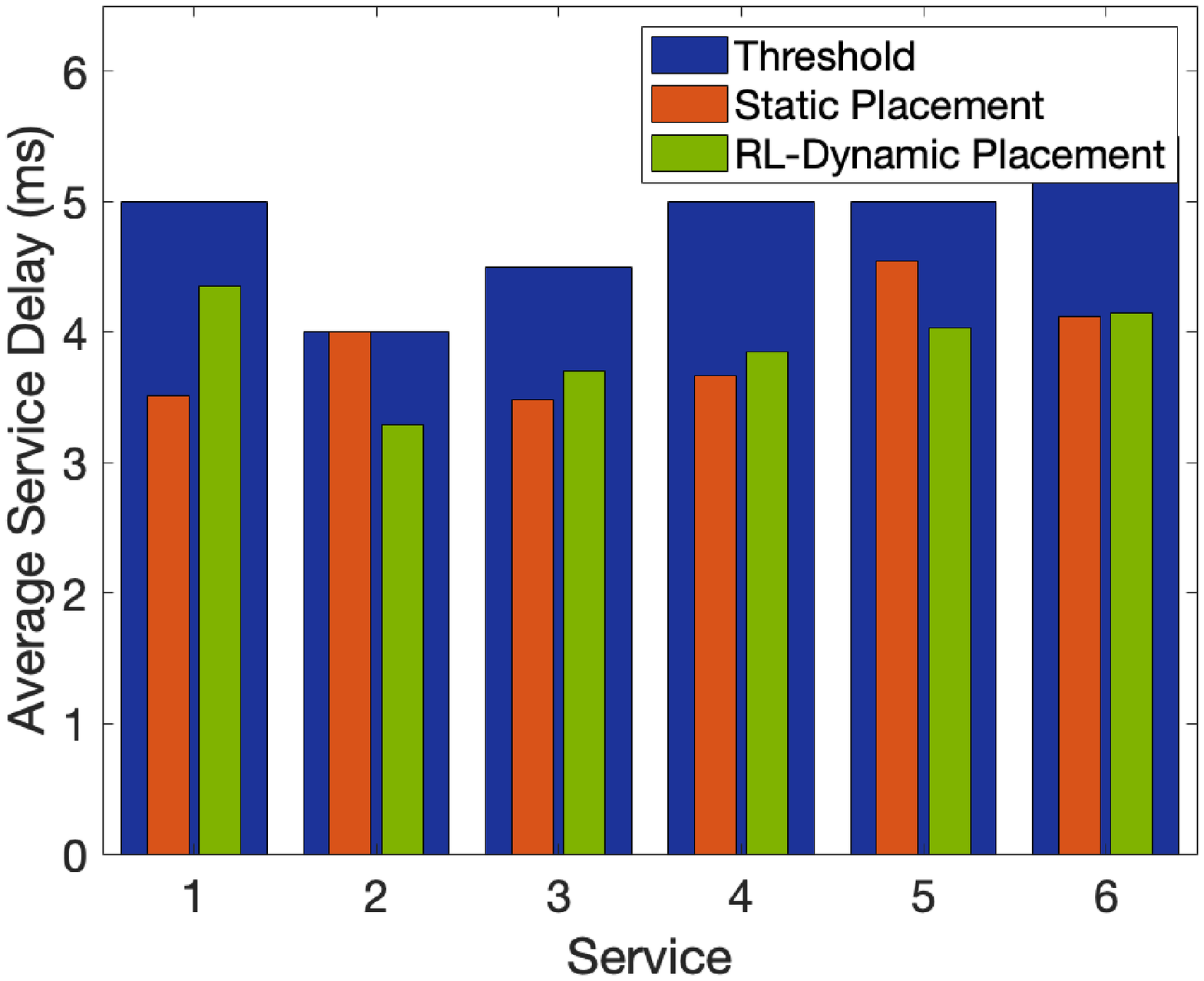}  
		\caption{SU-Opt (Trial 05)}
		\label{fig:delay10}
	\end{subfigure} 
	\caption{Average service delay for the two optimization algorithms.}
	\label{fig:compdelay}
\end{figure*}

Fig. \ref{fig:compdelay} shows the average delay experienced for different
services by the vehicles. It compares the average service delay
of the static placement and RL-Dynamic placement scenarios
(for both optimization models). Here, the static placement
refers to one-time solution of an ILP formulation. It can
be observed that the average delay observed by vehicles for
the static service placement is higher than the RL-Dynamic
service placement. This is due to the fact that the vehicular
environment is not stationary. The high mobility of nodes and
constantly changing topology requires continual learning of
the environment (as in the case with our RL-Dynamic service
placement) to provide a better average delay for each service.
It can be observed that the static placement is not able to
satisfy the delay requirement for some services. In contrast,
the delay for RL-Dynamic placement is always well below
its threshold for all trials. Moreover, with the SU-optimization
model, although the delay is not the lowest when compared
with the D-optimization model, it is always under the required
threshold value. This shown the effectiveness of adopting a
learning method by the dynamic approach. \par 

\subsection{Edge Server Utilization}
In this section, we consider edge server utilization and evaluate the performance in terms of fairness and average percentage of utilization. The fairness of server utilization are a representation of fair and efficient resource consumption. We use Jain's index as a fairness measure in this work \cite{jain_index}. The server utilization is fairer when Jain$'$s index is closer to 1. \par
\begin{figure}[htbp]
	\centering
	\begin{subfigure}{.21\textwidth}
		\centering
		\includegraphics[width=1.6in]{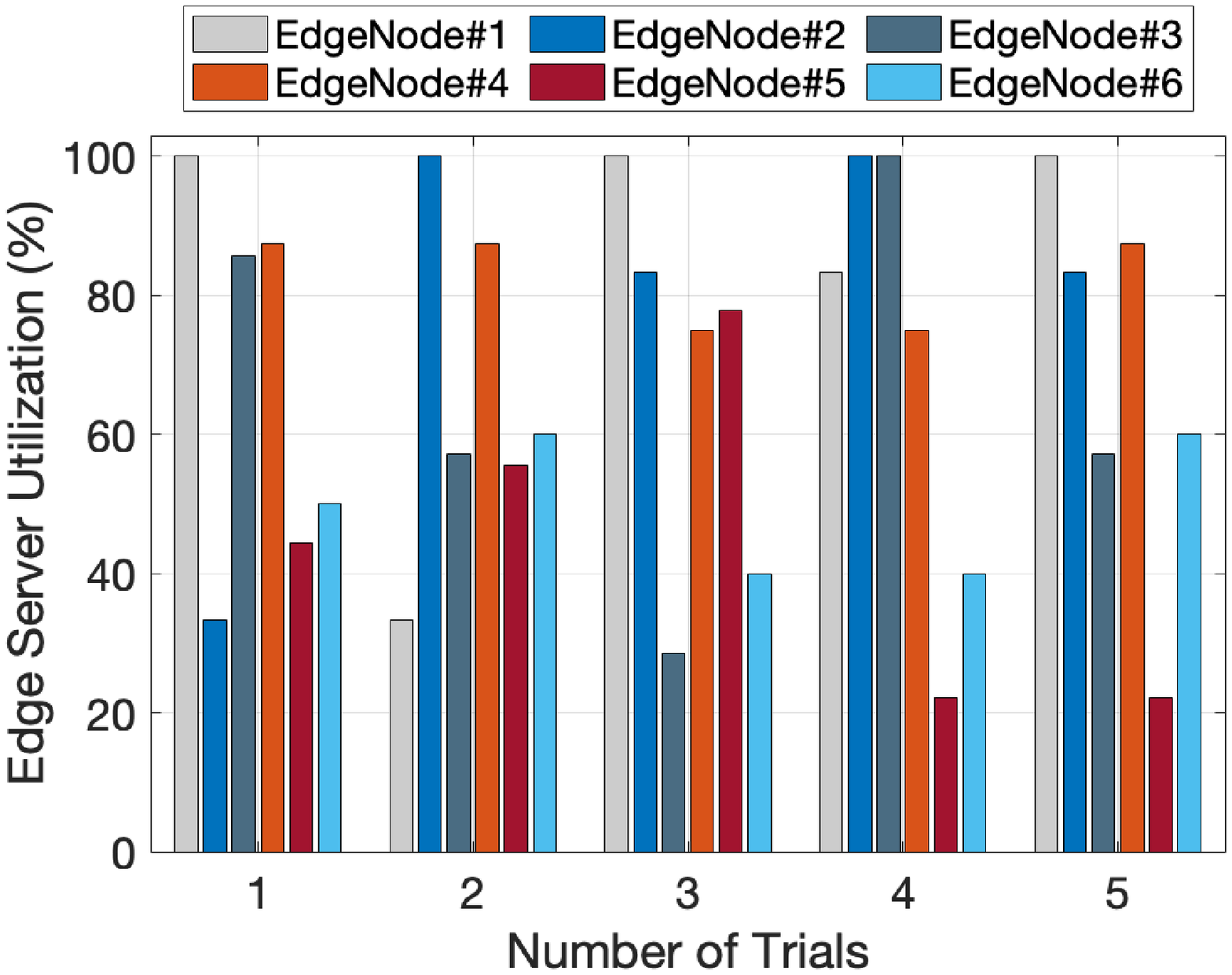}  
		\caption{Static (D-Opt)}
		\label{fig:su1}
	\end{subfigure}
	\begin{subfigure}{.21\textwidth}
		\centering
		\includegraphics[width=1.6in]{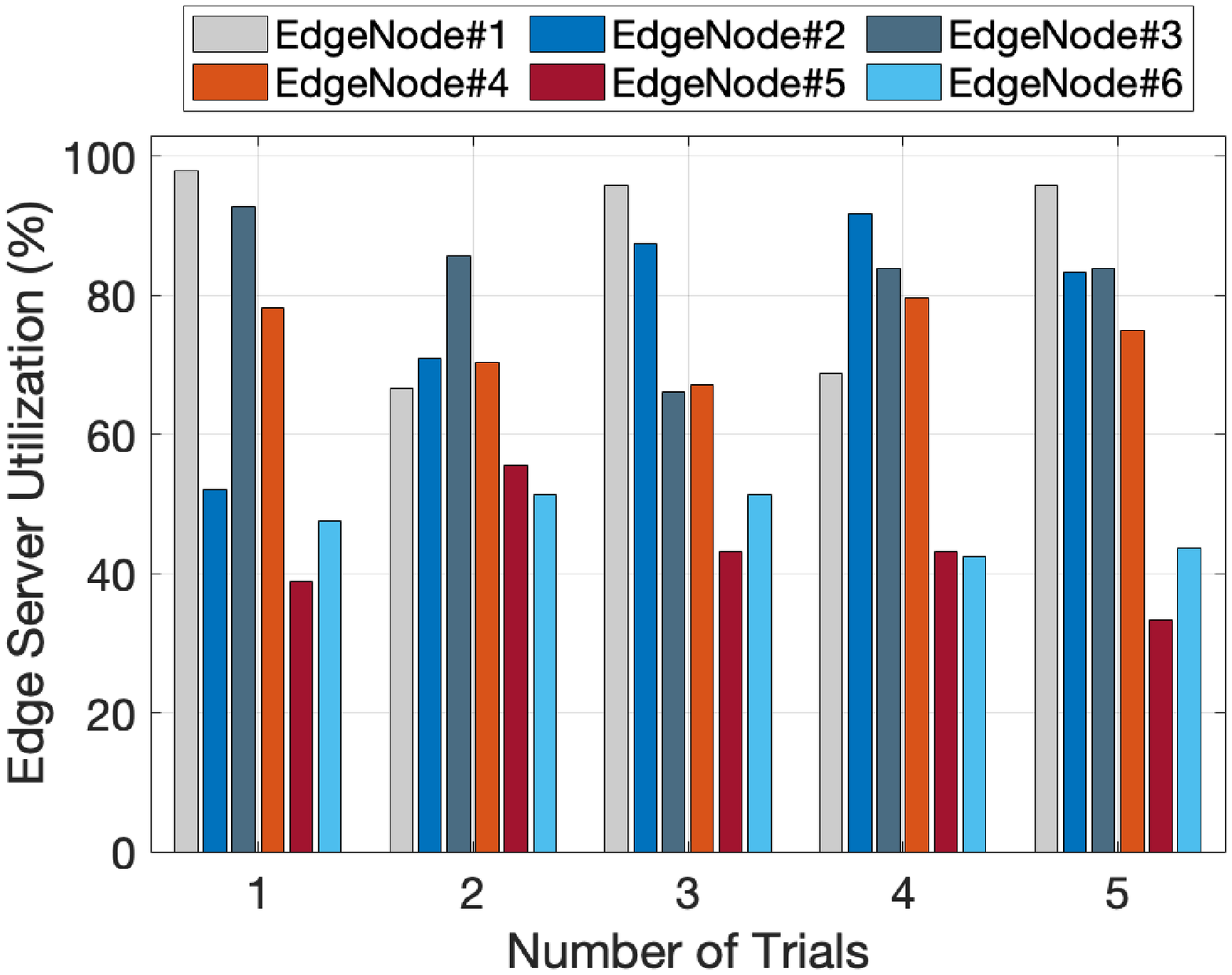}  
		\caption{RL-Dynamic (D-Opt)}
		\label{fig:su2}
	\end{subfigure}  \\
	\begin{subfigure}{.21\textwidth}
		\centering
		\includegraphics[width=1.6in]{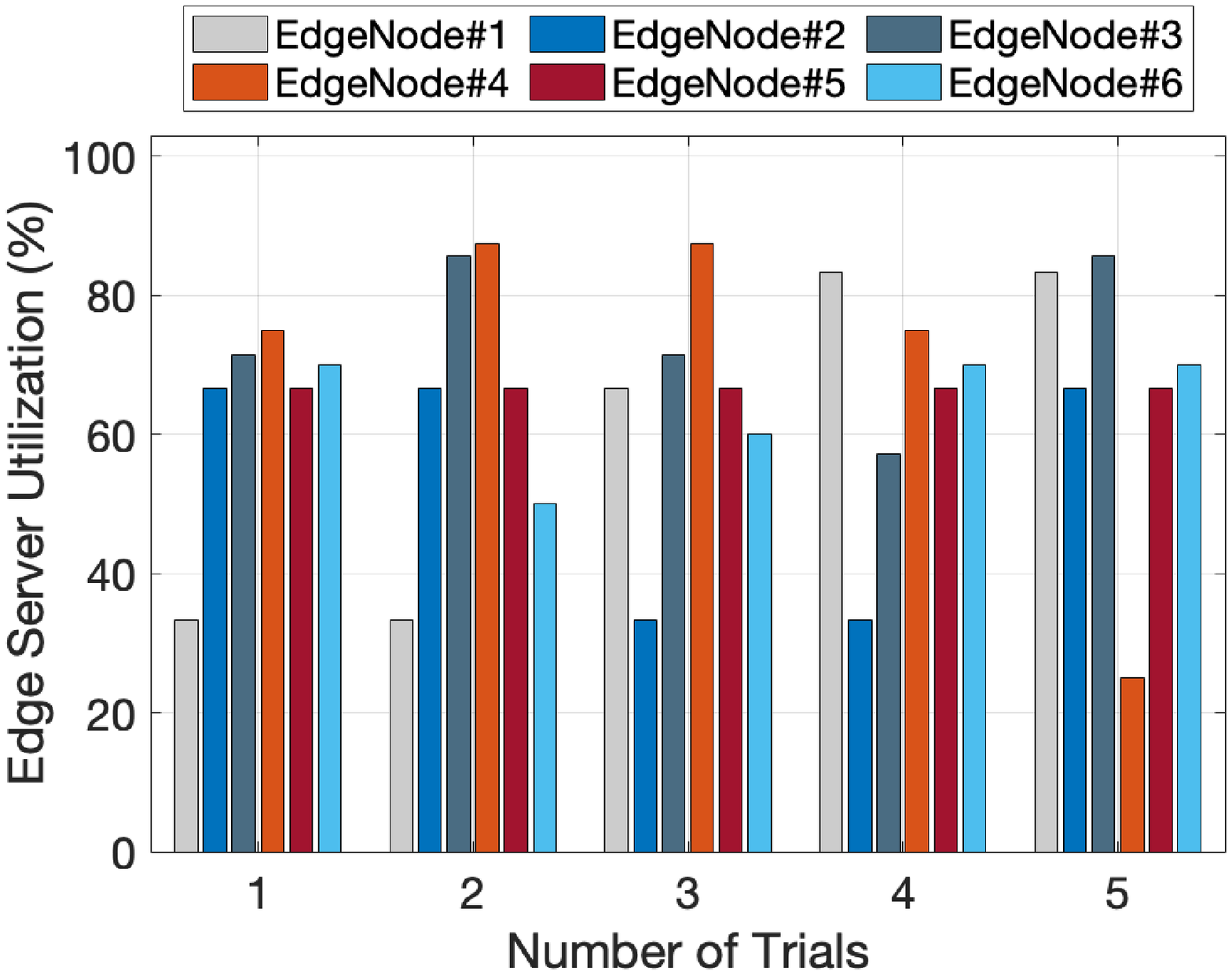}  
		\caption{Static (SU-Opt)}
		\label{fig:su3}
	\end{subfigure}
	\begin{subfigure}{.21\textwidth}
		\centering
		\includegraphics[width=1.6in]{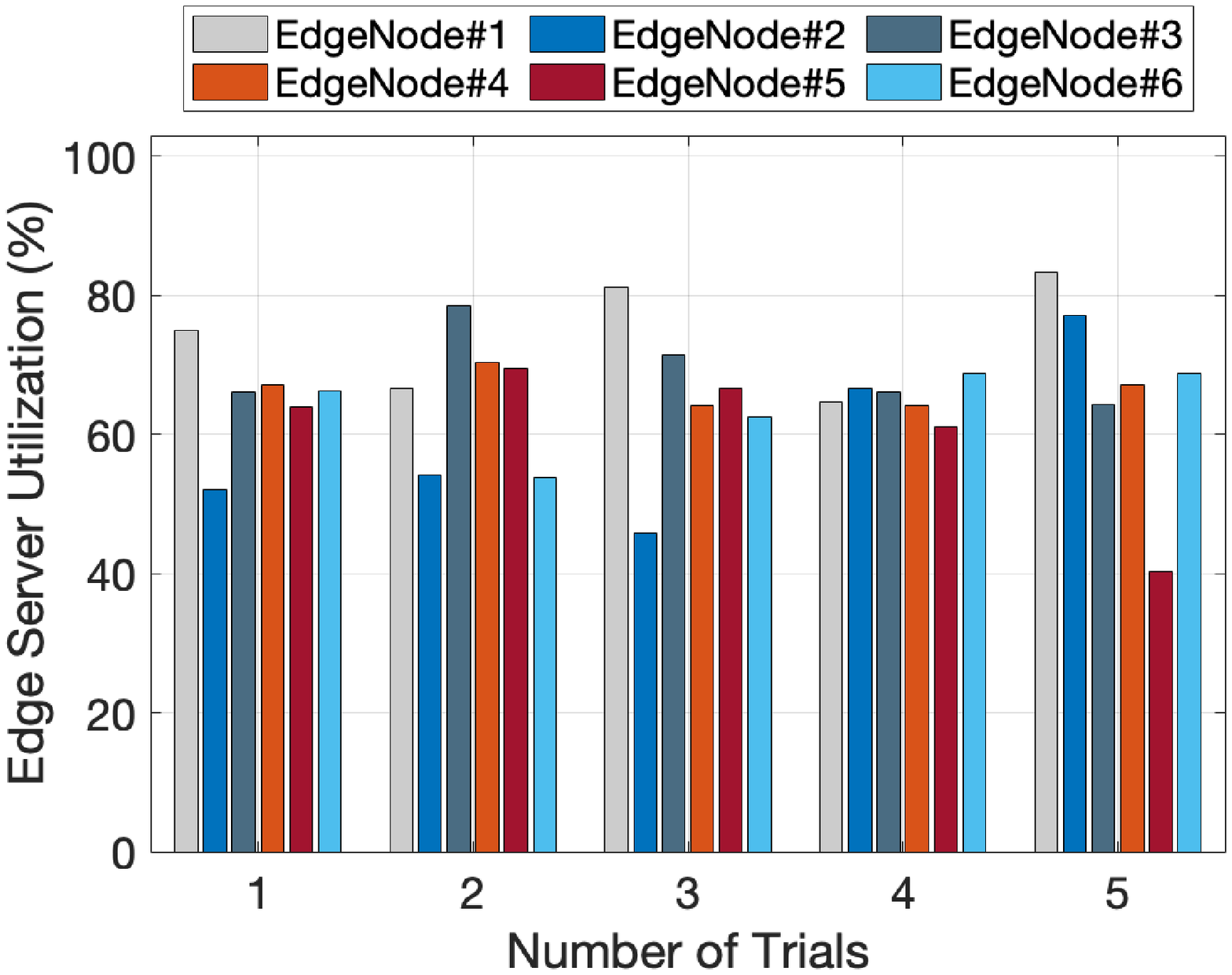}  
		\caption{RL-Dynamic (SU-Opt)}
		\label{fig:su4}
	\end{subfigure}
	\caption{Edge server utilization for the two optimization algorithms.}
	\label{fig:compsu}
\end{figure}

 
\begin{table}[htbp]
	\centering
	\caption{Fairness (Jain's index)}
	\begin{tabular}{|c|c|c|c|c|}
		\hline
		\multirow{2}[4]{*}{\textbf{Trial\#}} & \multicolumn{2}{c|}{\textbf{D-Optimization}} & \multicolumn{2}{c|}{\textbf{SU-Optimization}} \\
		\cline{2-5}        & \textbf{Static} & \textbf{RL-Dynamic} & \textbf{Static} & \textbf{RL-Dynamic} \\
		\hline
		1     & 0.876 & 0.898 & 0.954 & 0.989 \\
		\hline
		2     & 0.899 & 0.972 & 0.921 & 0.982 \\
		\hline
		3     & 0.879 & 0.932 & 0.940 & 0.974 \\
		\hline 
		4     & 0.850 & 0.926 & 0.942 & 0.999 \\
		\hline 
		5     & 0.877 & 0.903 & 0.917 & 0.961 \\
		\hline 
	\end{tabular}%
	\label{tab:fair_comp}%
\end{table}%
\begin{figure}[htbp]
	\centering
	\begin{subfigure}{.22\textwidth}
		\centering
		\includegraphics[width=1.6in]{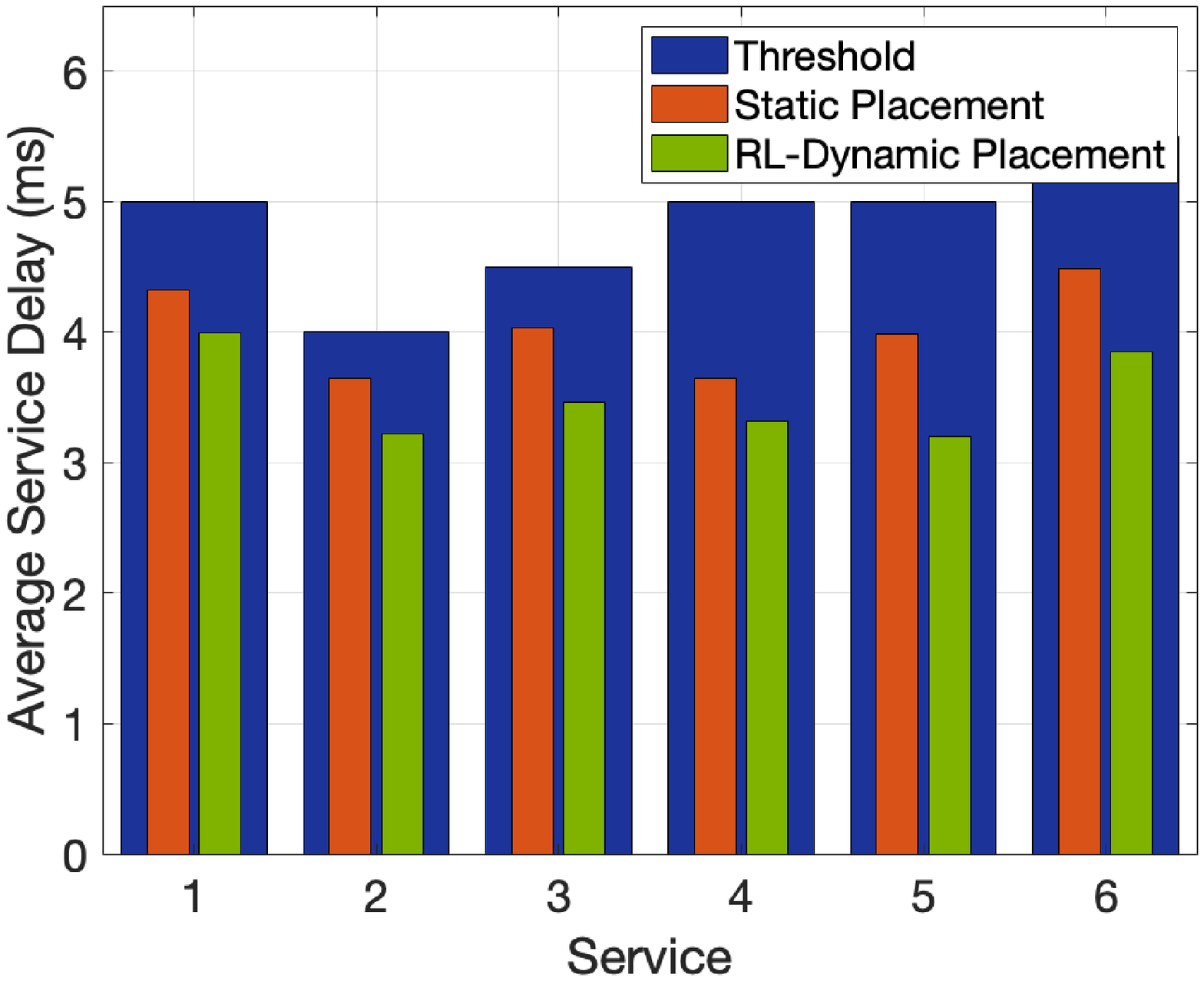}  
		\caption{D-Optimization}
		\label{fig:delayav1}
	\end{subfigure}
	\begin{subfigure}{.22\textwidth}
		\centering
		\includegraphics[width=1.6in]{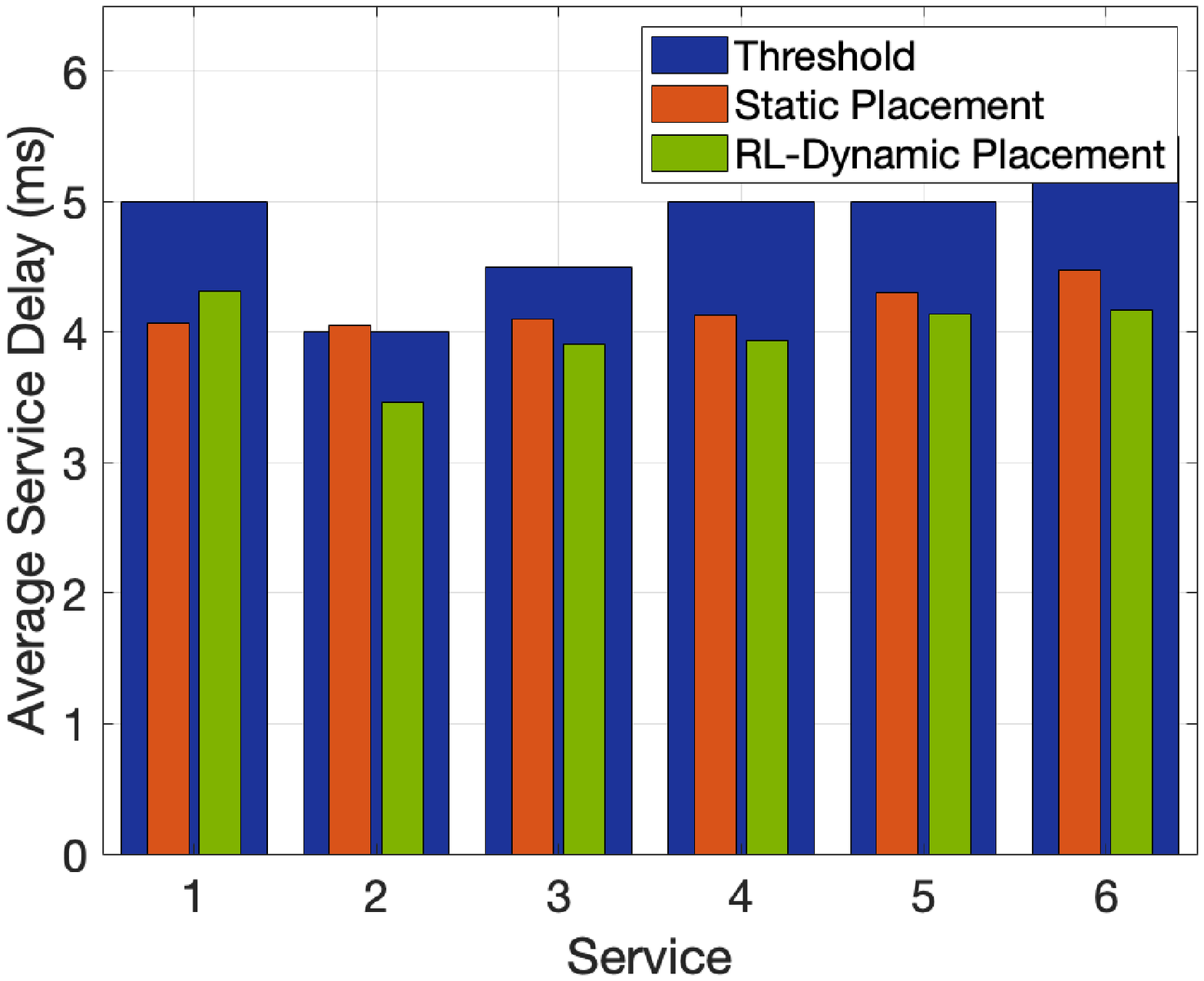}  
		\caption{SU-Optimization}
		\label{fig:delayav2}
	\end{subfigure} 
	\caption{Average service delay for the two optimization algorithms (Average of five trials).}
	\label{fig:compdelayavg}
\end{figure}
From Table. \ref{tab:fair_comp}, we can observe that the proposed scheme of
RL-Dynamic service placement achieves higher fair utilization
of edge servers compared to the static one. When compared
to the D-optimized, the SU-optimized model exhibits substantially
higher fairness in server utilization. In addition, as shown
in Fig. \ref{fig:compsu}, the proposed SU-optimized model mitigates the load
imbalance problem and spreads the load more evenly across
the edge nodes, while satisfying the delay requirements. The
balanced spread of service resources among different edge
nodes will also help to prevent the saturation/congestion at
any single server given the limited resources at the servers.
Inefficient usage of resources not only results in wastage but
also forces future service demands to be accessed from the
network core that will incur higher delay leading to lower
performance. \par
\begin{table}[htbp]
	\centering
	\caption{Average edge server utilization (\%)}
	\begin{tabular}{|c|c|c|c|c|}
		\hline
		\multirow{2}[4]{*}{\textbf{Trial\#}} & \multicolumn{2}{c|}{\textbf{D-Optimization}} & \multicolumn{2}{c|}{\textbf{SU-Optimization}} \\
		\cline{2-5}        & \textbf{Static} & \textbf{RL-Dynamic} & \textbf{Static} & \textbf{RL-Dynamic} \\
		\hline
		1     & 66.83 & 67.89 & 63.85 & 65.08 \\
		\hline
		2     & 65.59 & 66.72 & 64.98 & 65.48  \\
		\hline
		3     & 67.45 & 68.48 & 64.27 & 65.29 \\
		\hline 
		4     & 70.09 & 68.26 & 64.25 & 65.21 \\
		\hline 
		5     & 68.37 & 69.20 & 66.23 & 66.82 \\
		\hline 
	\end{tabular}%
	\label{tab:SU_comp}%
\end{table}%

Now, we compare the server utilization for different algorithms
and optimizations, as shown in Table. \ref{tab:SU_comp}. It is the
average of the percentage of the available capacity of all
servers consumed by a service. As can be observed from the
table, our proposed SU-optimized placement intends to utilize
edge-server resources more effectively accommodating the
same demand (as carried out by D-optimized), but with lower
percentage of edge server utilization. It is also worth noting
that we performed multiple trials to show the confidence level
of the performance.

\section{Conclusion}
\label{sec:conclude}
In this work, we addressed the problem of dynamic service
placement in vehicular networks. We developed a reinforcement
learning-based algorithm for continual learning of the environment to capture the dynamicity of vehicles and varying service demands and request-types. For the decision module in
our learning framework, we explored two different objective
functions- minimizing the delay and minimizing the server
resource utilization. We developed ILP problem formulations
for the two objective functions. We evaluated our framework
by simulating a virtual traffic scenario of a realistic vehicular
network using SUMO. Our performance study shows that
our proposed RL-based dynamic service placement achieves
higher fairness in utilization of edge server nodes and low
service delay compared to the static one. When compared
to D-optimized decision, the SU-optimized decision utilizes
resources more effectively balancing the load across edge
servers, and achieving higher fairness with lower edge server
utilization. As a future work, we plan to extend this work
considering attack scenarios and a more complex architecture
of edge networks.

\bibliographystyle{IEEEtran}
\bibliography{IEEEabrv,References}

\end{document}